\documentclass[conference]{IEEEtran}

\makeatletter
\def\ps@headings{%
\def\@oddhead{\mbox{}\scriptsize\rightmark \hfil \thepage}%
\def\@evenhead{\scriptsize\thepage \hfil \leftmark\mbox{}}%
\def\@oddfoot{}%
\def\@evenfoot{}}
\makeatother
\usepackage{flushend}

\usepackage{bigstrut,multirow,array,psfig,epsfig,subfigure,amssymb,amsmath,color,url,xr}
\usepackage[ruled,linesnumbered]{algorithm2e}

\begin{document}
%
\title{Dial It In: Rotating RF Sensors \\to Enhance Radio Tomography}

\author{
\IEEEauthorblockN{Maurizio Bocca, Anh Luong, Neal Patwari, and Thomas Schmid}
\IEEEauthorblockA{Electrical \& Computer Engineering Department\\
University of Utah, Salt Lake City, USA\\
maurizio.bocca@utah.edu, anh.n.luong@utah.edu, npatwari@ece.utah.edu, thomas.schmid@utah.edu}
}

\maketitle

\begin{abstract}
A radio tomographic imaging (RTI) system uses the received signal strength (RSS) measured by RF sensors in a static wireless network to localize people in the deployment area, without having them to carry or wear an electronic device.  This paper addresses the fact that small-scale changes in the position and orientation of the antenna of each RF sensor can dramatically affect imaging and localization performance of an RTI system. 
However, the best placement for a sensor is unknown at the time of deployment.  Improving performance in a deployed RTI system requires the deployer to iteratively ``guess-and-retest'', \emph{i.e.}, pick a sensor to move and then re-run a calibration experiment to determine if the localization performance had improved or degraded. We present an RTI system of \emph{servo-nodes}, RF sensors equipped with servo motors which autonomously ``dial it in'', \emph{i.e.}, change position and orientation to optimize the RSS on links of the network. By doing so, the localization accuracy of the RTI system is quickly improved, without requiring any calibration experiment from the deployer. Experiments conducted in three indoor environments demonstrate that the servo-nodes system reduces localization error on average by 32\% compared to a standard RTI system composed of static RF sensors.
\end{abstract}




%
\IEEEpeerreviewmaketitle

\section{Introduction}
\label{sec:introduction}

Radio tomographic imaging (RTI) systems \cite{Wilson_RTI_2010,Patwari_2010_IEEE} localize and track people in indoor areas using the received signal strength (RSS) measurements made by a network of multiple static wireless devices.  These devices are called ``RF sensors'' because their RF interface is their mode of sensing. Instead of requiring people to carry an electronic device (\emph{e.g.}, RFID tag, mobile phone, etc.), an RTI system uses the changes in RSS on the network's links to estimate the attenuation field caused by the presence and movements of people found in it. RTI systems can be used to enable context awareness \cite{Grandma_2012,Xu_2013,MTT_2013,Nannuru_2013}, in ambient-assisted living (AAL) applications \cite{EvAAL_book,falling_detection}, and in tactical operations or crisis situations \cite{Timonen_RTI,Joint_UWB_RSS}.

\begin{figure}[t]
    \begin{center}
        \epsfig{figure=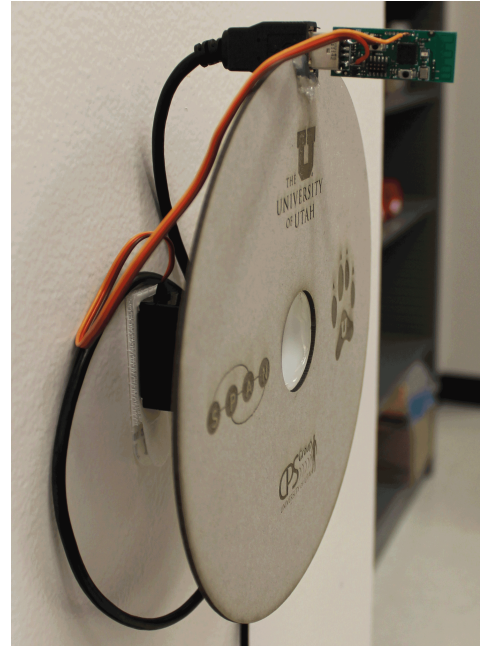,width=0.7\columnwidth}
        \caption{Servo-node Platform. The automated prototype used in this work is composed of a TI CC2531 RF sensor, operating in the $2.4$ GHz ISM band, and a servo motor that can rotate one full turn ($360$ degrees). The RF sensor is glued on a rigid cardboard disc having a $10$ cm radius. The disc is in turn glued on the winch of the servo motor. The sensor controls the position of the servo motor through one of its I/O ports.}
        \label{fig:servo_node}
    \end{center}
\end{figure}

An effect we have observed over many deployments is that the performance of an RTI system can be dramatically altered (improved or degraded) by small (sub-wavelength) position changes of the deployed RF sensors.  Two RTI deployments in the same area, with RF sensors deployed in ostensibly the same positions, may have significantly different tracking performance.  We show an example of how RTI performance is improved by moving one sensor in Section \ref{sec:fade_level_effect}.

One may systematically improve RF sensor locations, and thus RTI system performance, by a long and tedius procedure we call ``guess-and-retest''.  First, RF sensors are deployed, and a experiment is conducted with the deployer moving in a known path, by which the tracking error of the RTI system is calculated.  Next, the deployer:
\begin{enumerate}
\item picks a sensor to be the \emph{sensor-under-test} and moves it a few cm in one direction or another.
\item re-performs the known-path experiment and re-calculates the error.  
\item If the tracking error increases, the deployer moves the sensor-under-test back to its original position.
\item Repeat from Step 1.
\end{enumerate}
While ``guess-and-retest'' is possible and will ultimately reduce localization error, it is extraordinarily time-consuming and as such is unsuitable for a real-world deployment of a commercial RTI system.  

In this paper, we introduce an RTI system composed of a network of the autonomously rotating RF sensors, which we call \emph{servo-nodes}, shown in Figure \ref{fig:servo_node}. Each servo-node is equipped with a servo motor and is capable of performing small-scale, \emph{i.e.}, on the order of a wavelength, adjustments of the position of the RF sensor.  Further, we suggest and justify a simple network-wide quality metric which is based solely on link channel measurements when no person is in the area, and thus does not require the deployer to conduct any known-path experiment.  Together, the quality metric and the servo-nodes allow each sensor to quickly, \emph{i.e.}, within seconds, ``dial it in'', \emph{i.e.}, rotate to optimize its own position.  We show via three deployments that this procedure, which we refer to as \emph{calibration}, reduces localization error by 30\% to 37\%.

The key to the improvement is in the optimization of link \emph{fade level}, the degree to which RSS is changed by constructive or destructive multipath fading.  If multipath components arrive at the receiver antenna with nearly the same phase, the link is said to be in \emph{anti-fade}, and its RSS is relatively high. Alternatively, if components have nearly opposite phase, a link is said to be in \emph{deep fade}, and its RSS is relatively low \cite{Wilson_SkewL_2011}. Since the phases of each component changes at a different rate as the antenna is moved, we observe the effect of small-scale fading \cite{Rappaport_book}.  

Previous works \cite{EvAAL_book,Wilson_SkewL_2011,MASS_2012,multi_scale_arxiv} have demonstrated that the change in RSS induced by a person obstructing the link line, \emph{i.e.}, the straight line connecting transmitter and receiver, strongly depends on the fade level of the link. Anti-fade links measure a consistent attenuation only when the person is located in the proximities of the link line. In contrast, deep fade links measure a variation in RSS (either increase or decrease) also when the person is located at unpredictable positions far away from the link line.  Anti-fade links thus provide generally more informative and reliable information about a person's position.

We propose that maximizing the sum of RSS on all links measured during empty-room conditions will increase, on average, link fade levels, and thus improve RTI tracking accuracy.  Servo-nodes don't move the sensors far enough to alter large scale path loss on links, thus any increase in RSS can be attributed to a change in small-scale fading that makes the multipath phasor sum more constructive.  With the link multipath arranged to be more constructive, there is a higher probability that the link will exhibit a more reliable and predictable attenuation behavior when obstructed by a person.  The predictable behavior of the link thus improves RTI localization accuracy. Instead of ``guess-and-retest'', the deployer simply turn on and deploys servo-nodes and leaves the room.  The sensors self-calibrate and rotate to a (local) optimum position.  Even better, the servo-nodes could periodically recalibrate to adjust to changing environmental conditions over months and years.

We present results from three deployments, \emph{i.e.}, a typical one bedroom apartment, a highly cluttered university laboratory, and a large office space. Preliminary experiments are conducted in the apartment with a multi-node platform (see Figure \ref{fig:cardboard_nodes} and Section \ref{sec:multi_node_platform}) that simulates the functioning of a servo-node. The servo-nodes (see Figure \ref{fig:servo_node} and Section \ref{sec:servo_nodes}) are used in the subsequent (lab and office) deployments. We also describe two different calibration procedures that iteratively adjust position and orientation of the nodes composing the rotating RTI system. Both procedures aim at increasing the overall RSS of the links of the network in static conditions, so as to create more anti-fade links and consequently improve the localization accuracy. The results of the deployments show that a system composed of rotating RF sensors in random positions, \emph{i.e.}, with random orientation, achieves a localization accuracy similar to the one of a \emph{standard} RTI system composed of static nodes all with the same orientation. However, when the servo-nodes are calibrated, RTI localization error is reduced on average by 30\% compared to a standard RTI system with the same number of sensors.  Alternatively, the calibrated servo-node system can achieve the same accuracy as a system of standard sensors, but with 37\% fewer sensors.





\section{Methods}
\label{sec:methods}

In this section, we describe the multi-node platform and the rotating servo-nodes used in the deployments, the procedures applied to calibrate the position and orientation of the RF sensors, and the RTI method used to process the RSS measurements collected in the experiments.

\subsection{Hardware}
\label{sec:hardware}

\subsubsection{Multi-node Platform}
\label{sec:multi_node_platform}

\begin{figure}[t]
    \begin{center}
        \epsfig{figure=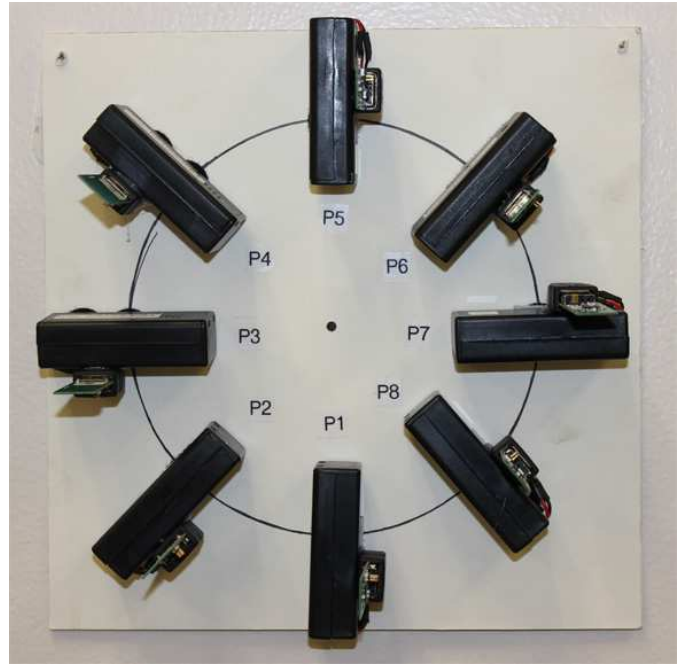,width=\columnwidth}
        \caption{Multi-node Platform. A prototype is built with eight battery-powered RF sensors positioned clockwise along the perimeter of a circle having a $10$ cm radius. Each sensor has different orientation.}
        \label{fig:cardboard_nodes}
    \end{center}
\end{figure}

To conduct preliminary experiments, we created a multi-node platform (see Figure \ref{fig:cardboard_nodes}) with eight battery-powered RF sensors attached to a rigid cardboard tile and positioned clockwise along the perimeter of a circle having a $10$ cm radius. Each sensor of the platform has different orientation. The platform was designed to simulate the functioning of the rotating servo-nodes (see Section \ref{sec:servo_nodes}), having an RF sensor in each of the eight positions where the servo motor can position its own sensor.

\subsubsection{Servo-nodes}
\label{sec:servo_nodes}

The servo-nodes are composed of two parts: the RF sensor, \emph{i.e.}, a TI CC2531 USB dongle \cite{tidongle}, and the servo motor, \emph{i.e.}, a GWS digital sail winch servo \cite{GWS125_servo}. The RF sensor operates in the $2.4$ GHz ISM band. It has a maximum nominal transmit power of $4.5$ dBm and can transmit on one of $16$ selectable frequency channels, which are $5$ MHz apart, as specified by the IEEE 802.15.4 standard. The servo motor can rotate one full turn ($360$ degrees) through the standard $1$-$2$ ms pulse width modulation (PWM). A rigid cardboard circle, having a $10$ cm radius, is glued to the winch of the servo motor. The RF sensor is in turn attached to the cardboard circle so that its antenna is perpendicular to the surface of the circle. The CC2531 platform controls the position of the servo motor through one of its I/O ports. We programmed the nodes so to be able to rotate them to eight different positions ($p = \{1,...,8\}$), $45$ degrees apart.

The RF sensors collect RSS measurements on the selected frequency channels by running the \emph{multi-Spin} communication protocol \cite{EvAAL_book}. The packets broadcasted by the servo-nodes and received at the central sink node include the RSS measurements of the links of the network and indicate the current position of the servo-nodes. multi-Spin reserves one slot at the end of each TDMA communication cycle in order for the sink node to communicate a new position to one of the servo-nodes.

\subsection{RF Sensors Calibration Procedures}
\label{sec:calibration_procedures}

We now introduce two different procedures to calibrate the small-scale position and orientation of the RF sensors. The first, which we refer to as \emph{incremental} calibration, was used with the multi-node platform for preliminary experiments carried out in the one bedroom apartment. The second, which we refer to as \emph{network} calibration, was used with the servo-nodes in experiments carried out in the laboratory and office space.

\subsubsection{Incremental Calibration}
\label{sec:incremental_calibration}

After positioning the first sensor (\emph{i.e.}, \#1 in Figure \ref{fig:apartment_nodes}), the other sensors are deployed and calibrated by applying the following iterative procedure:
\begin{enumerate}
    \item Pick a spot to temporarily deploy the multi-node platform. The spot is chosen so as to maximize the length of the calibrated links and cover the whole deployment area uniformly (\emph{e.g.}, one can iteratively use the four cardinal points as a reference);
    \item In static conditions, \emph{i.e.}, with no people in the deployment area, measure for a short period of time (\emph{e.g.}, 10 s) the RSS of all the links among the eight sensors on the multi-node platform and the sensors already calibrated and deployed.
    \item For sensor $p \in \{1,...,8\}$ on the multi-node platform, calculate $\bar{R}_p$, \emph{i.e.}, the mean of the time-averaged RSS in static conditions of all the links between $p$ and the other sensors $d \in \mathcal{D}$ already calibrated and deployed, as:
        \begin{equation} \label{eq:incremental_calibration}
            \bar{R}_p = \frac{1}{|\mathcal{D}|} \frac{1}{|\mathcal{C}|} \sum_{d \in \mathcal{D}} \sum_{c \in \mathcal{C}} (\bar{r}_{(p,d),c} + \bar{r}_{(d,p),c}),
        \end{equation}
    where $\mathcal{C}$ is the set of measured frequency channels and $|C|$ its cardinality.
    \item Remove the multi-node platform and permanently deploy a sensor exactly where the sensor $p$ with the highest value of $\bar{R}_p$ was. Add the newly deployed sensor to the set $\mathcal{D}$.
    \item Repeat step 1) through 4).
\end{enumerate}

\subsubsection{Network Calibration}
\label{sec:network_calibration}

In the incremental calibration, the RF sensors are deployed and calibrated one at a time. In the network calibration, first, all the RF sensors composing the system are deployed. With the servo motors in their default position ($p=1$),  we measure for a short period of time (\emph{e.g.}, $10$ s) the RSS of all the $L$ links of the network in static conditions. At the end of this period, we calculate $\bar{R}$ as the mean of the time-averaged RSS in static conditions of the $L$ links of the network on the frequency channels in $\mathcal{C}$:
    \begin{equation} \label{eq:mean_RSS_network}
        \bar{R}= \frac{1}{L} \frac{1}{|\mathcal{C}|}  \sum_{l=1}^L \sum_{c \in \mathcal{C}} \bar{r}_{l,c}.
    \end{equation}
Starting from sensor \#1, we apply to each sensor $s$ of the system the following calibration procedure:
\begin{enumerate}
    \item Collect $M$ RSS measurements ($M = 10$ in our tests) for each link and frequency channel, \emph{i.e.}, collect RSS measurements for the time interval corresponding to $M$ multi-Spin TDMA communication cycles.
    \item Calculate and store $\bar{R}_s^p$, \emph{i.e.}, $\bar{R}$ with the servo motor of sensor $s$ in position $p$.
    \item While $p < 8$, rotate sensor $s$ to the next position and repeat steps 1) and 2).
    \item If $\max{(\bar{R}_s^p)} > \bar{R}_N$, rotate sensor $s$ to the corresponding position $p$ and set $\bar{R} = \max{(\bar{R}_s^p)}$. Otherwise, rotate sensor $s$ back to its original position, \emph{i.e.}, the position sensor $s$ had at the end of the last iteration of the calibration procedure.
\end{enumerate}
The calibration procedure is repeated until the rotating RF sensors do not set into different positions compared to the previous iteration, \emph{i.e.}, until $\bar{R}$ does not increase anymore.

\subsection{Radio Tomographic Imaging}
\label{sec:RTI}

\begin{figure*}[t]
    \begin{center}
        \mbox{
            \subfigure[\quad Simulation \emph{A}]{\epsfig{figure=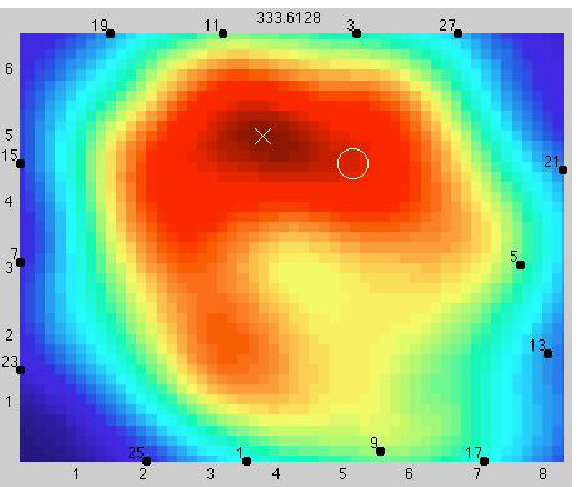,width=\columnwidth}} \quad
            \subfigure[\quad Simulation \emph{B}]{\epsfig{figure=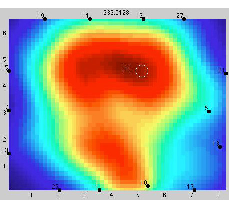,width=\columnwidth}}
        }
        \caption{The effect of sensor's position. RTI images formed by an RTI system composed $14$ standard sensors deployed in a $54$ m$^2$ highly cluttered laboratory at the University of Utah. In simulation \emph{A}, sensor \#9 is selected, and the localization error is $1.01$ m. In simulation \emph{B}, sensor \#10 is selected, and the localization error is $0.52$ m. The two sensors are $20$ cm apart. In the images, the white circle represents the true position of the person, the white cross the estimated position.}
        \label{fig:RTI_comparison}
    \end{center}
\end{figure*}

In this section, we summarize the RTI method, introduced in \cite{multi_scale_arxiv}, used to process the RSS measurements collected in the experiments. To the best of our knowledge, the method in \cite{multi_scale_arxiv} is, to date, the RTI method achieving the highest localization accuracy. By processing the data with this method, we aim at demonstrating that a system composed of rotating RF sensors can further enhance the localization accuracy of RTI. While in this work we consider only the method in \cite{multi_scale_arxiv}, we expect similar improvements in localization accuracy by using rotating RF sensors also with other RTI methods \cite{Wilson_VRTI_2011,MASS_2012,Zhao_2013}.

An RTI system is composed of $N$ RF sensors deployed at known positions $\{x_n,y_n\}_{n=1,...,N}$ and communicating on a set $\mathcal{C}$ of different frequency channels. At each time instant $k$, the system measures the RSS $r_{l,c}(k)$ of link $l$ on frequency channel $c \in \mathcal{C}$. By combining the RSS measurements collected on all the $L = N \cdot (N-1)$ links of the network on the $C$ selected frequency channels, the system estimates in real-time the change in the propagation field of the monitored area caused by people found in it.

During an initial training phase of the system, performed when the deployment area is not occupied by people, we measure the average RSS of each link on each measured frequency channel. We denote this as $\bar{r}_{l,c}$. After the training phase, we estimate the RSS attenuation of link $l$ on channel $c$ at time instant $k$ as:
\begin{equation} \label{eq:link_attenuation}
    \Delta{r}_{l,c}(k) = r_{l,c}(k)-\bar{r}_{l,c}.
\end{equation}
In RTI, the attenuation field to be estimated is discretized into voxels. The attenuation of a link is assumed to be a spatial integral of the RF propagation field of the monitored area. Thus, for each link, the change in RSS is a linear combination of the change in the attenuation of a subset of voxels, \emph{i.e.}, the voxels within an ellipse having the transmitter and receiver of the link at the foci.

While in previous works the width of the ellipse $\lambda$ was set to a fixed value for all the links of the network, the method in \cite{multi_scale_arxiv} defines a different value $\lambda_{l}$ for each link based on its \emph{fade level} \cite{Wilson_SkewL_2011}. The fade level of link $l$ on channel $c$ is defined as:
\begin{equation} \label{eq:fade_level}
    F_{l,c} = \bar{r}_{l,c} - P(d,c),
\end{equation}
where $P(d,c)$ is the theoretical RSS estimated by using the log-distance path loss model \cite{Rappaport_book}, which depends on the distance $d$ between transmitter and receiver and on the center frequency $c$. The path loss exponent $\eta$ of the model is derived after the initial calibration by applying linear least squares fitting to the measured mean RSS of all the links of the network. After this, the fade level of the links on each selected frequency is calculated as in (\ref{eq:fade_level}).

As defined in \cite{Wilson_SkewL_2011}, an \emph{anti-fade} link-channel pair $(l,c)$ has positive fade level, while a \emph{deep fade} one has negative fade level. The characteristics of these two types of links have been described in \cite{Wilson_SkewL_2011} and then modeled in \cite{multi_scale_arxiv}. The work in \cite{Wilson_SkewL_2011} demonstrated that the sensitivity area of deep fade links is larger than the one of anti-fade links. In addition, when a deep fade link is obstructed, on average the measured RSS increases. Instead, when an anti-fade link is obstructed, on average the measured RSS decreases. The model described in \cite{multi_scale_arxiv} determines  for each link-channel pair $(l,c)$ two parameters, $\lambda^{+}$ and $\lambda^{-}$, \emph{i.e.}, the width of the ellipse for a measured increase $(+)$ and decrease $(-)$ in RSS, respectively, based on the fade level. The value of $\lambda^{-}$ is considerably smaller for anti-fade links than for deep fade links, as anti-fade links measure a decrease in RSS only when the person is located in the close proximity of the link line. Thus, anti-fade links provide higher quality information for the purpose of device-free localization. By calibrating the position of the servo-nodes, our system increases the mean RSS of the links of the network, pushing the links towards an anti-fade like behavior, which improves the localization accuracy.

For each link-channel pair, based on the so determined $\lambda^{+}_{l,c}$ and $\lambda^{-}_{l,c}$, its fade level, and the magnitude of the change in RSS calculated as in (\ref{eq:link_attenuation}), the model in \cite{multi_scale_arxiv} derives the probabilities $p^{+}_{l,c}$ (for an increase in RSS) and $p^{-}_{l,c}$ (for a decrease in RSS) of the person being within the area defined by the corresponding ellipse: the larger the change in RSS, the higher the probability.

When all the $L$ links of the network are considered, the changes in the propagation field of the monitored area can be estimated as:
\begin{equation}\label{eq:linear_formulation}
    \mathbf{y} =  \mathbf{W}\mathbf{x}+\mathbf{n},
\end{equation}
where $\mathbf{y}$ and $\mathbf{n}$ are the measurements and noise vectors and $\mathbf{x}$ is the image to be estimated. The measurements vector is composed of the probabilities
$p^{+}_{l,c}$ and $p^{-}_{l,c}$ for each link-channel pair of the network. The elements of the weight matrix $\mathbf{W}$, representing how the attentuation of voxel $j$ affects the RSS measurements of the links, are calculated as:
\begin{equation}\label{eq:weight_matrix}
w^{\delta}_{l,c,j} = \begin{cases}
        \frac{1}{A^{\delta}_{l,c}} & \text{if } d_{l,j}^{tx}+d_{l,j}^{rx}<d_{l}+\lambda^{\delta}_{l,c}\\
        0 & \text{otherwise}
    \end{cases},
\end{equation}
where $\delta = \{+,-\}$ represents the sign of the change in RSS, $A^{\delta}_{l,c}$ is the area of the ellipse, $d_{l,j}^{tx}$ and $d_{l,j}^{rx}$ are the distances from the center of voxel $j$ to the transmitter and receiver, respectively, and $d_{l}$ is the length of the link.

Since the number of links $L$ is considerably smaller than the number of voxels of the image, the estimation problem is an ill-posed one, and regularization has to be applied. We use a regularized least-squares approach \cite{Patwari_corr_shadow_2008}:
\begin{equation}\label{eq:regularization}
    \hat{\mathbf{x}} =  \mathbf{\Pi}\mathbf{y}.
\end{equation}
The projection matrix $\mathbf{\Pi}$ is calculated as:
\begin{equation}\label{eq:projection_matrix}
    {\mathbf{\Pi}} = {(\mathbf{W}^T\mathbf{W}+\mathbf{C}_{x}^{-1}\sigma_{N}^{2})}^{-1}\mathbf{W}^T,
\end{equation}
where $\sigma_{N}^{2}$ is the regularization parameter. The \emph{a priori} covariance matrix $\mathbf{C}_{x}$ is calculated by using an exponential spatial decay:
\begin{equation}\label{E:cov_matrix}
    [\mathbf{C}_{x}]_{j,i}=\sigma_{x}^{2}e^{-d_{j,i} /\delta_{c}},
\end{equation}
where $\sigma_{x}^{2}$ is the variance of voxel measurements, $d_{j,i}$ is the distance between the center of voxel $j$ and the center of voxel $i$, and $\delta_{c}$ is the voxels' correlation distance. The position of one person located in the monitored area is estimated by finding the coordinates of the voxel of the RTI image that has the maximum value.

\begin{figure}[t]
    \begin{center}
        \epsfig{figure=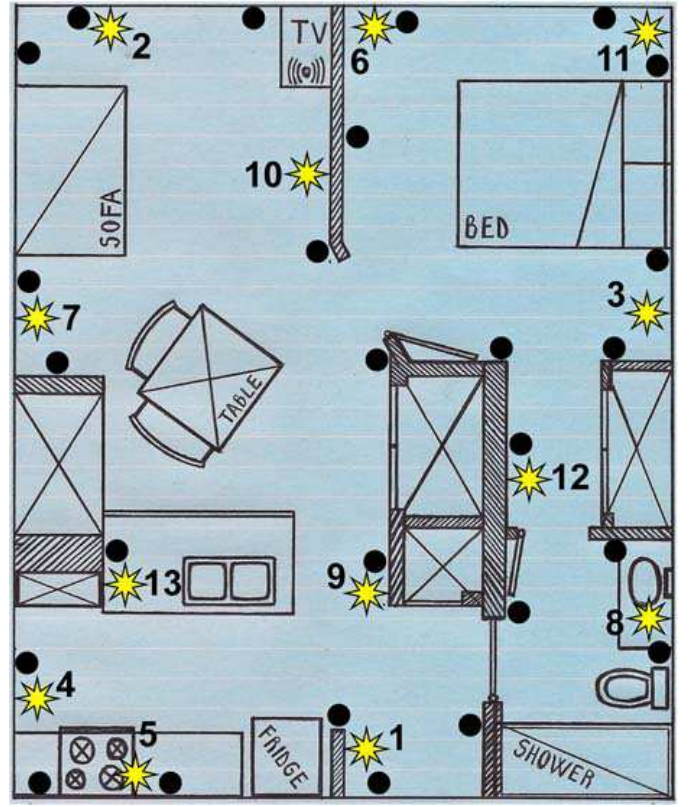,width=\columnwidth}
        \caption{Floor map of the one bedroom apartment used in the experiments. The yellow stars represent the RF sensors calibrated with the multi-node platform. The black dots represent the standard RF sensors, all having the same orientation.}
        \label{fig:apartment_nodes}
    \end{center}
\end{figure}

\section{Experimental Results}
\label{sec:experimental_results}

We now present results from three different deployments of our system. In each deployment, we compare the performance of the new servo-nodes system to a system composed of \emph{standard} RF sensors, \emph{i.e.}, static sensors all having the same orientation and running the multi-Spin communication protocol. To make a fair comparison of the performance of the two systems, two standard RF sensors are positioned in the proximity of each servo-node, one on each side, at distance $d \le 20$ cm from the winch of the servo motor. In this way, we are ensuring that the links among the nodes of both systems cover the deployment area approximately in the same way, and that the differences in accuracy do not originate from the nodes being positioned at different locations.

\subsection{Effect of Sensors Position on RTI}
\label{sec:fade_level_effect}

First, we show how small-scale changes in the position of the RF sensors affect the imaging and localization performance of an RTI system. To do this, we perform two different simulations by using the same RSS measurements collected with the standard nodes during a test in a $54$ m$^2$ highly cluttered laboratory at the University of Utah. For both simulations (\emph{A} and \emph{B}), we select $14$ of the $28$ deployed standard RF sensors (\emph{i.e.}, one for each of the $14$ deployed servo-nodes). $13$ of the selected $14$ standard sensors are the same in both simulations. However, in simulation \emph{A}, we select sensor \#9, while in simulation \emph{B} we select sensor \#10. These two sensors are $20$ cm apart. Figure \ref{fig:RTI_comparison} shows the RTI images formed in simulation \emph{A} and \emph{B} when the person is located at coordinates $(5.13,4.57)$ m. The localization error of simulation \emph{A} is $1.01$ m, while the error of simulation \emph{B} is $0.52$ m.

\begin{figure}[t]
    \begin{center}
        \epsfig{figure=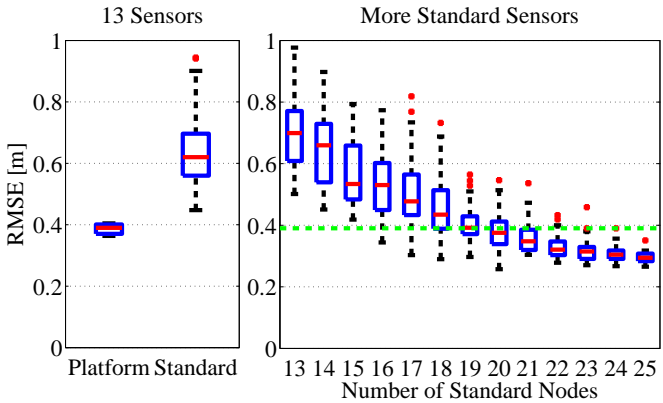,width=\columnwidth}
        \caption{Results of the experiments in a $56$ m$^2$ one bedroom apartment. On the left, the RMSE of the system composed of $13$ sensors calibrated with the multi-node platform is compared to the RMSE measured with different subsets of $13$ standard sensors. On the right, the RMSE measured with a varying number of standard nodes. The horizontal line represents the RMSE measured with $13$ sensors calibrated with the multi-node platform.}
        \label{fig:apt_218_RMSE_results}
    \end{center}
\end{figure}

\begin{figure*}[t]
    \begin{center}
        \mbox{
            \subfigure[\quad Laboratory at the University of Utah]{\epsfig{figure=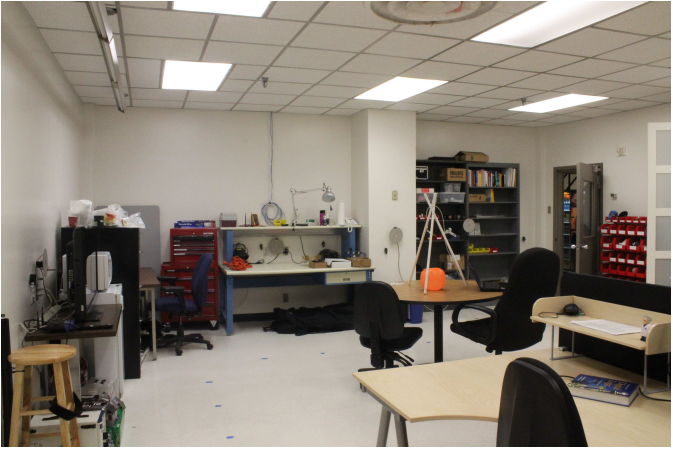,width=\columnwidth}} \quad
            \subfigure[\quad Office space at the University of Utah]{\epsfig{figure=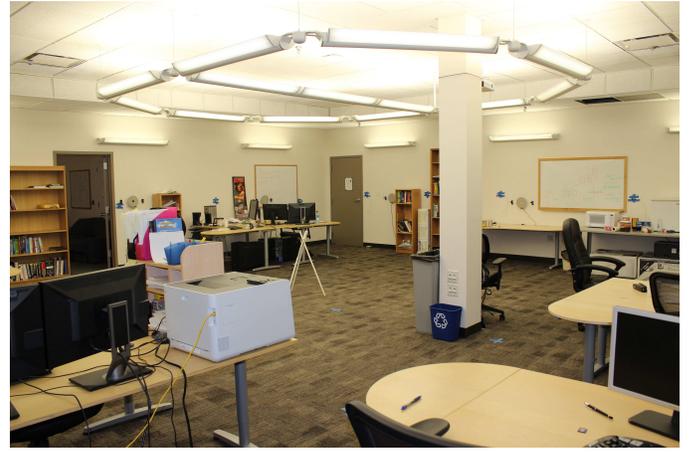,width=\columnwidth}}
        }
        \caption{Servo-nodes deployments: in (a), the $54$ m$^2$ laboratory. In (b), the $100$ m$^2$ office space.}
        \label{fig:deployment_environments}
    \end{center}
\end{figure*}

\subsection{Preliminary Experiments}
\label{sec:one_bedroom_apartment_deployment}

Preliminary experiments were conducted in a $56$ m$^2$ one bedroom apartment (with floor map shown in Figure \ref{fig:apartment_nodes}). We deployed and calibrated $13$ RF sensors by using the multi-node platform described in Section \ref{sec:multi_node_platform} and the incremental procedure described in Section \ref{sec:incremental_calibration}. We also deployed $26$ standard sensors, \emph{i.e.}, sensors having the same orientation, communicating on the same frequency channels ($\mathcal{C} = \{11,16,21,26\}$). For the standard sensors, we chose positions in the proximities of the spots selected for the sensors calibrated with the multi-node platform. All the RF sensors were deployed at approximately $1.2$ m from the floor.

To evaluate the localization accuracy, we marked $45$ points on the floor of the apartment. These points represented the true position of the person to be localized during a test. We asked the person to stand still at each of these locations for $8$ s. In this work, we consider only the accuracy in localizing a stationary person in order to provide a more reliable comparison of the performance of the rotating and standard RTI systems. However, the methods and systems presented in this work can be used to localize and track a moving person.

Figure \ref{fig:apt_218_RMSE_results} shows the box plot of the results of five different tests conducted in the one bedroom apartment. In each test, the person stood in the $45$ evaluation points with different orientations of her body. With the $13$ sensors calibrated with the multi-node platform, the median root mean squared error (RMSE) is $0.39$ m. For each test, we create $30$ different subsets of $13$ standard sensors by selecting, for each calibrated sensor, one of the two neighboring standard sensors. In this way, we ensure that the links connecting the selected sensors cover the deployment area uniformly. The simulations are performed by using the data collected during the tests. The median RMSE of the resulting $150$ simulations is $0.62$ m. Thus, the system composed of calibrated RF sensors reduces the localization error by $37\%$ compared to a standard system composed of sensors all having the same orientation.

We also perform simulations by increasing the number of sensors composing the standard RTI system. For each test, we create  $30$ different subsets of standard sensors. In this case, the sensors are chosen randomly. The results of the simulations show that the median RMSE decreases with a larger number of sensors. With $19$ standard sensors, the median RMSE is $0.39$ m, \emph{i.e.}, the same as the median RMSE obtained with the calibrated sensors. Thus, by calibrating the sensors with the multi-node platform, we are able to achieve the same localization accuracy by using $32\%$ fewer sensors.

\subsection{Servo-nodes Deployments}
\label{sec:servo_nodes_deployments}

We now describe the results of two deployments in which we used the servo-nodes described in Section \ref{sec:servo_nodes} and the network calibration procedure described in Section \ref{sec:network_calibration}.

The servo-nodes were first deployed in a $54$ m$^2$ highly cluttered laboratory at the University of Utah (see Figure \ref{fig:deployment_environments}(a)). We deployed a total of $14$ servo-nodes. We also deployed two standard RF sensors in the proximity of each servo-node, one on each side, at a $20$ cm distance from the winch of the servo motor. In this deployment, the set of used frequency channels $\mathcal{C} = \{15,20,25,26\}$, in order to minimize the interference with multiple coexisting WiFi networks.

To evaluate the localization accuracy, we marked $32$ points on the floor. The points were chosen so as to cover all the areas of the laboratory. First, we carried out $10$ tests by having the $14$ servo-nodes positioned in $10$ different permutations of random positions. The mean RMSE of these tests was $0.59$ m. Then, for each test, we created $10$ different subsets of $14$ standard sensors by selecting, for each servo-node, one of the two neighboring standard sensors. The mean RMSE of the resulting $100$ simulations was $0.61$ m. The simulations were performed by using the same data collected during the tests. These results demonstrate that the RTI system composed of servo-nodes, when these are not calibrated, achieves on average a localization accuracy very similar to the one achieved by a standard RTI system.

\begin{figure}[t]
    \begin{center}
        \epsfig{figure=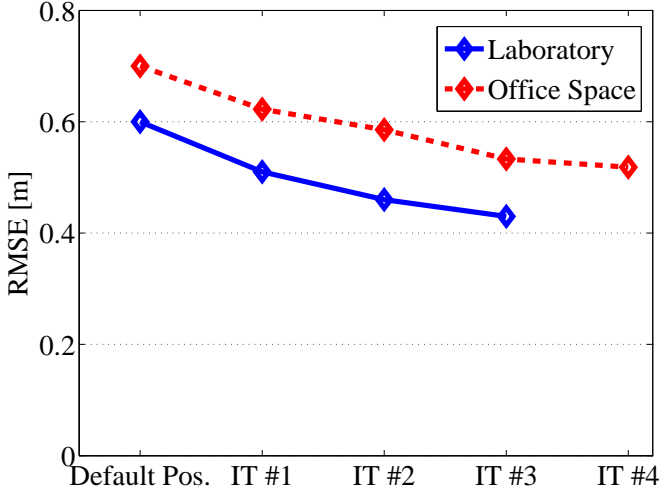,width=\columnwidth}
        \caption{Results of the two servo-nodes deployments. In both deployments, the RMSE decreases at each iteration of the network calibration procedure. In the laboratory (blue line), the RMSE with the servo-nodes in default position ($p = 1$) is $0.60$ m. After the third and final iteration, the RMSE is $0.43$ m. In the office space (red dashed line), the RMSE with the servo-nodes in default position is $0.70$ m. After the fourth and final iteration, the RMSE is $0.52$ m. By calibrating the RF sensors, we achieve a $30\%$ reduction of the localization error in both deployments.}
        \label{fig:servo_nodes_results}
    \end{center}
\end{figure}

Subsequently, we applied the network calibration procedure to the servo-nodes. Figure \ref{fig:servo_nodes_results} shows the mean RMSE of three different tests performed after each iteration of the calibration procedure. After the first iteration, the mean RMSE is $0.51$ m, \emph{i.e.}, a $16\%$ reduction of the localization error achieved with the standard sensors. After the second iteration, the mean RMSE is $0.46$ m, \emph{i.e.}, a $25\%$ reduction. After the third and final iteration, the mean RMSE is $0.43$ m, \emph{i.e.}, a $30\%$ reduction.

The last deployment was carried out in a $100$ m$^2$ office space at the University of Utah (see Figure \ref{fig:deployment_environments}(b)). We deployed a total of $12$ servo-nodes ($3$ on each side of the space) and $24$ standard sensors (two in the proximities of each servo-node, one per side). The nodes density of this deployment ($0.12$ nodes/m$^2$) was considerably lower than in other previous works that use RTI methods and low-power sensors operating in the $2.4$ GHz ISM band: for example, in \cite{multi_scale_arxiv,Wilson_SkewL_2011,MASS_2012,Grandma_2012,MTT_2013}, the nodes density assumed values at least four times higher. In this deployment, the nodes communicated on frequency channels $\mathcal{C} = \{15,20,25,26\}$.

To evaluate the localization accuracy of both RTI systems, we marked $23$ points on the floor of the office space. First, we carried out $8$ tests by having the $12$ servo-nodes positioned in $8$ different permutations of random positions. The mean RMSE of these tests was $0.72$ m. For each test, we created $10$ different subsets of $12$ standard sensors by selecting, for each servo-node, one of the two neighboring standard sensors. The mean RMSE of the resulting $80$ simulations was $0.74$ m. The simulations were performed by using the data collected during the tests. Also in this deployment, the system composed of servo-nodes in random positions and the system composed of standard sensors have very similar localization accuracy.

Figure \ref{fig:servo_nodes_results} shows the mean RMSE of three different tests performed after each iteration of the network calibration procedure. The RMSE is $0.62$ m after the first iteration, \emph{i.e.}, a $16\%$ reduction of the localization error achieved with the standard sensors, $0.59$ m after the second iteration, \emph{i.e.}, a $20\%$ reduction, $0.53$ m after the third iteration, \emph{i.e.}, a $28\%$ reduction, and $0.52$ m after the fourth and final iteration, \emph{i.e.}, a $30\%$ reduction.

\begin{figure}
    \begin{center}
        \epsfig{figure=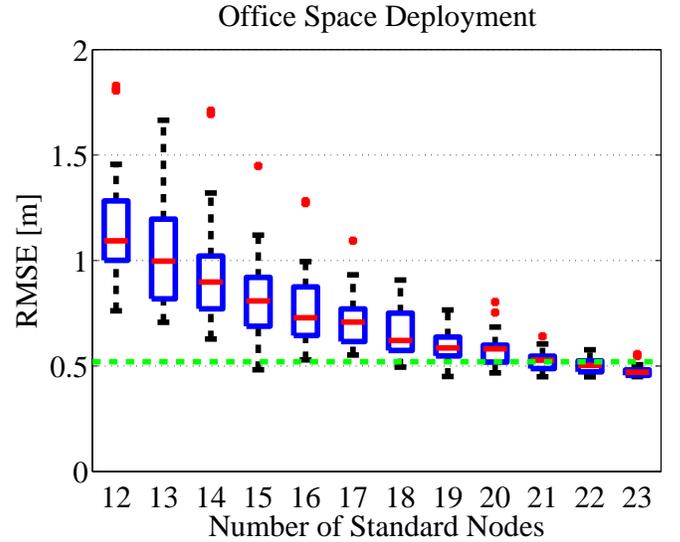,width=\columnwidth}
        \caption{RMSE measured with a varying number of standard sensors in a $100$ m$^2$ office space at the University of Utah. The horizontal dashed line represents the RMSE measured with $12$ servo-nodes after the fourth and final iteration of the calibration procedure.}
        \label{fig:large_room_sims_more_nodes}
    \end{center}
\end{figure}

Figure \ref{fig:large_room_sims_more_nodes} shows the results of simulations performed by increasing the number of sensors composing the standard RTI system. We create $150$ different subsets of standard sensors for each number of standard nodes. The sensors are chosen randomly. The median RMSE decreases with a higher number of sensors. With $21$ standard sensors, the mean RMSE is $0.52$ m, \emph{i.e.}, the same mean RMSE measured with $12$ calibrated servo-nodes. Thus, the RTI system composed of servo-nodes achieves the same accuracy by using $43\%$ fewer sensors. The results of the three deployments are summarized in Table \ref{T:results_summary}.



\begin{figure}
    \begin{center}
        \epsfig{figure=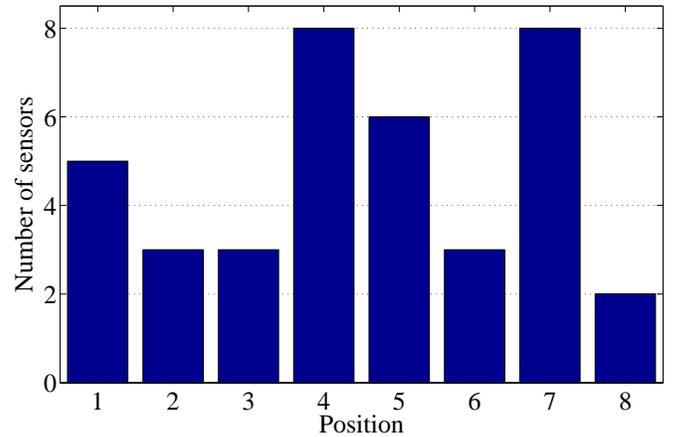,width=\columnwidth}
        \caption{Distribution of the calibrated positions of the RF sensors in the three deployments.}
        \label{fig:sensors_pos_distr}
    \end{center}
\end{figure}

\begin{center}
\begin{table*}[t!]
    \caption{Summary of the Results} 
        \centering
        \footnotesize
        \begin{tabular}{c  c  c | c  c  c  c | c}
        \hline\hline 
        \\ [-2.0ex]
        &  &  & \multicolumn{3}{c}{RMSE [m]} & \\
        \\ [-2.0ex]
        \hline
        \\ [-2.0ex]
        Deployment & Area [m$^2$] & \# of Nodes & \shortstack{Standard\\Sensors} & \shortstack{Servo-nodes\\Random Pos.} & \shortstack{Servo-nodes\\Default Pos.} & \shortstack{Calibrated\\Positions} & Improvement\\
        \\ [-2.0ex]
        \hline
        \\ [-2.0ex]
        Apartment    & $56$  & $13$ & $0.62$ &        &        & $0.39$ & $37\%$\\ 
        \\ [-2.0ex]
        Laboratory   & $54$  & $14$ & $0.61$ & $0.59$ & $0.60$ & $0.43$ & $30\%$\\ 
        \\ [-2.0ex]
        Office space & $100$ & $12$ & $0.74$ & $0.72$ & $0.70$ & $0.52$ & $30\%$\\ 
        \\ [-2.0ex]
        \hline 
         \multicolumn{5}{c}{} & \multicolumn{2}{r|}{\bf Average:} & \bf 32\% \\
        \end{tabular}
        \label{T:results_summary}
\end{table*}
\end{center}

\subsection{Position Distribution}

Does the calibrated position, \emph{i.e.}, the position at which the servo-nodes come to rest at after the calibration procedure, have a bias to one direction or another?  If so, we might suspect that a certain orientation is better because of the antenna polarization, or that a certain orientation results in a beneficial antenna gain pattern.  Alternatively, if the highest position (\#5 for our servo-nodes) ends up being chosen more often, we might suspect that the increase in antenna height is really to credit for the improvements in RTI performance. 

To address these questions, we plot the histogram of calibrated positions across all three deployments. In Figure \ref{fig:sensors_pos_distr}, we consider the calibrated positions of the multi-node platform used in the one bedroom apartment, and the final positions of the servo-nodes in the laboratory (\emph{i.e.}, after the third iteration) and office space (\emph{i.e.}, after the fourth iteration). There are a total of $38$ calibrated positions since we have $13$, $14$, and $12$ nodes in the apartment, laboratory, and office space, respectively, and the first node in the apartment is fixed.  

There does not seem to be any particular bias in any direction.  The maximum in the histogram is $8$, at positions $4$ and $7$.  If it is true that the eight positions are equally likely (our null hypothesis), using the multinomial distribution, we find that $86.9\%$ of the time, at least one of the positions would have $9$ occurrences or more.  Thus having $8$ occurrences of a particular position is not evidence to reject the equally likely positions hypothesis, and thus, we see no position bias.



\section{Related Work}
\label{sec:related_work}

Over the last few years, RTI has quickly become one of the most popular techniques of device-free localization. However, other RSS-based techniques have also proven to be feasible and accurate. The work in \cite{youssef07} introduced device-free passive (DfP) localization, which leverages typical wireless data network deployments and off-the-shelf wireless cards. This technique has been used to localize and track multiple people in cluttered \cite{SPOT_Youssef} and in large \cite{Nuzzer_Youssef} indoor environments.

Fingerprinting methods have been used in \cite{Xu_2013,Xu_2012} to estimate in which \emph{cells} people are located. The work in \cite{Xu_2012} uses probabilistic methods based on discriminant analysis. However, these methods require a long calibration period, \emph{i.e.}, $15$ to $30$ minutes. Other works have specifically tackled this issue, aiming at creating a localization system that would not require an extensive training phase to be carried out in static conditions. The work in \cite{Rabbat_background} uses background subtraction methods typical of machine vision to estimate baseline RSS values from measurements collected while the system is already in use and people may be located in the monitored area. In the context of RTI, the work in \cite{Grandma_2012} applied a low-pass filter to the RSS measurements of the links of the network in order to adapt the baseline RSS to the changes in the environment and make the system able to provide accurate position estimates in the long-run in a domestic environment.

Other systems use the time-of-flight (ToF) of radio signals to perform the localization task. The work in \cite{ToF_RTI} exploits the fact that, similar to the RSS, also the ToF of radio signals is affected by a person obstructing the link line. Thus, ToF measurements are used to form RTI images and localize the person in the monitored area. The system in \cite{MIT_3D_tracking} consists of a single device with one antenna for transmission and three for reception. The device transmits a radio signal and then measures the ToF of the signals reflected by the person's body. A geometric reference model is then used to map the ToF measurements of the receiving antennas to the position of the person.



\section{Conclusion}
\label{sec:conclusion}

The small-scale position of RF sensors significantly affects the performance of an RTI system.  A ``good'' position for each sensor is not known \emph{a priori}.  We provide an automated deployment system with servo-controlled RF sensors which rotate in a $10$ cm radius.  Each sensor in the network rotates, iteratively, and the system increases the average RSS measured on links network-wide.  By so doing, the system increases the average link fade level and improves radio tomographic image estimates.  This automated position calibration, which we refer to as ``dialing it in'', does not require any known-path experimentation from the deployer and occurs within seconds.  We demonstrate the system in three experimental deployments and show that it can reduce localization error by over $30\%$ compared to na\"ive sensor placement. In future work, we will consider using a platform with multiple, electrically switchable antennas, such as the prototype in \cite{SPIDA_antenna}, in order to increase the quality of the RSS measurements collected by each deployed RF sensor.






\end{document}